# Spin-valve Effect in Nanoscale Si-based Devices

Duong Dinh Hiep[1], Masaaki Tanaka[2,3] and Pham Nam Hai[1,3]

[1] Department of Electrical and Electronic Engineering, Tokyo Institute of Technology
[2] Department of Electrical Engineering and Information Systems, University of Tokyo
[3] Center for Spintronics Research Network, Graduate School of Engineering, University of Tokyo

**ABSTRACT**

The silicon (Si) based spin-MOSFET (metal-oxide-semiconductor field-effect transistor) is considered to be the building block of low-power-consumption electronics, utilizing spin-degrees of freedom in semiconductor devices. In this paper, we review the latest results on the spin transport in nanoscale Si-based spin-valve devices, which is important to realize the nanoscale spin-MOSFET. Our results demonstrate the importance of ballistic transport in obtaining high spin-dependent output voltage in nanoscale Si spin-valve devices.

**INTRODUCTION**

Present computing technology is based on the manipulation of electron charge currents. However, this electronic charge-based computing technology faces serious problems of high idling power consumption and heat generation due to the leakage off-current when a device's size is miniaturized, which restricts a device's operating speed. Hence, there is demand for alternative low-power solutions to overcome these problems in the beyond-CMOS (complementary metal-oxide-semiconductor) era. The silicon (Si) based spin-MOSFET can be a promising solution because of its high compatibility with the well-established CMOS technology and long spin lifetime in Si [1,2,3,4,5]. For this reason, there has been a great interest in demonstrating spin injection and detection of spin transport in Si by ferromagnetic electrodes. In fact, spin injection into microns of Si channels by using the three terminal Hanle effect [6,7,8] or the four terminal spin-valve effect [9,10,11] has been demonstrated. However, previous studies reported the spin transport only in micron-scale Si channels and the typical spin-dependent output voltages were less than 1 mV, which is not enough for realistic spin-MOSFET applications.

To improve the spin-dependent output voltage, we have proposed and demonstrated the spin-valve effect utilizing ballistic electron transport in nanoscale Si channels. It is expected that the ballistic transport of electrons in such nanoscale channels may overcome the conductivity mismatch problem that arises at the interface between a ferromagnetic (FM) electrode and a diffusive semiconductor (SC) channel [12,13], resulting in higher spin-dependent output voltage. In our studies, we fabricated Si-based spin-valve devices with a ferromagnetic spin injector / detector and a Si channel as short as 20 nm and sys-tematically investigated their spin-dependent transport characteristics [14,15]. We have observed a clear spin-valve effect up to 3% and a spin output voltage up to ~ 20 mV, which is the highest value reported so far in lateral spin-valve devices [15]. We also found that the sign of the spin-valve effect is reversed at low temperatures, suggesting the possibility of the spin-blockade effect of defect states in the MgO/Ge tunneling barrier [15].

**DEVICE FABRICATION**

In this study, we used a highly doped n-type Si (100) substrate with an electron density $n = 1 \times 10^{18}$ cm$^{-3}$. The Si substrates were first cleaned by an H$_2$SO$_4$/H$_2$O$_2$ solu-tion, then etched by diluted hydrofluoric acid solution to remove the native oxide layer, and rinsed in de-ionized water. To form a tunneling barrier and ferromagnetic electrodes for our devices (see Fig. 1(a)), we used two different deposition methods; electron-beam (EB) evaporation and molecular-beam epitaxy (MBE).

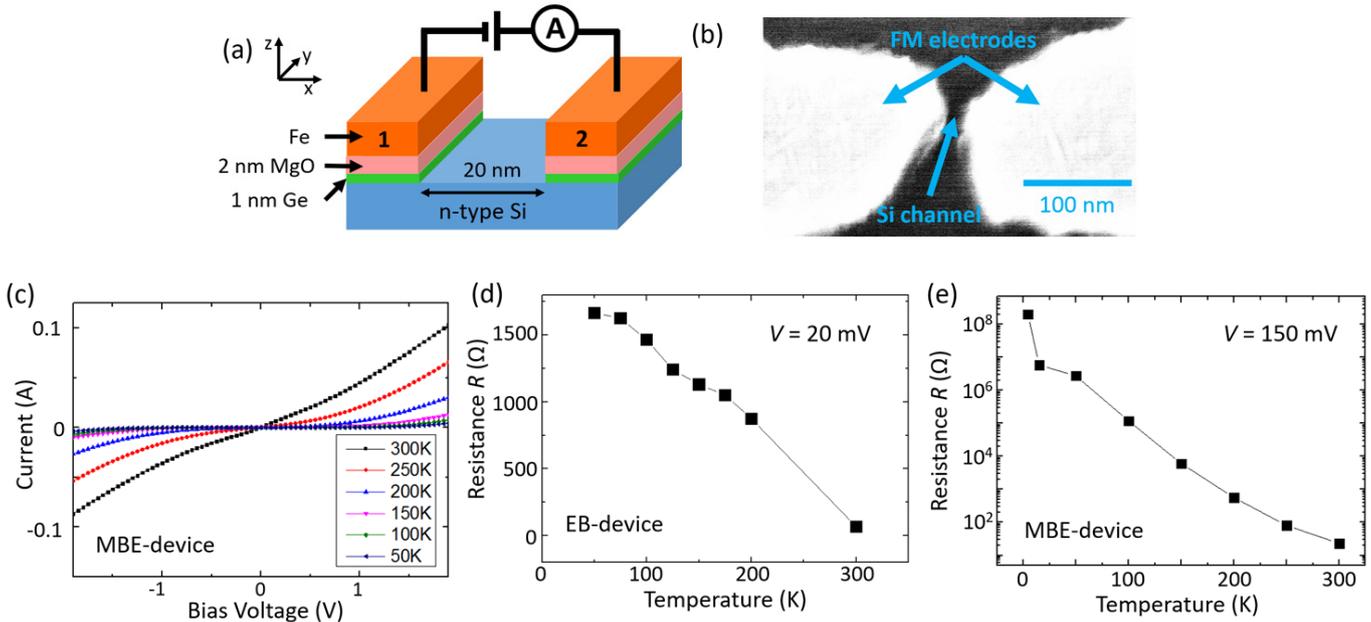

**Fig. 1:** (a) Schematic spin-valve device structure with an Fe/MgO/Ge spin injector / detector and a 20 nm-long Si channel. (b) Scanning electron microscopy image (top view) of a device. A silicon channel with a length of 20 nm was formed between the Fe electrodes. (c) Current–voltage characteristics (*I-V* curves) of a MBE-device at various temperatures. (d)-(e) Temperature dependence of the resistance of a EB-device measured at 20 mV and the MBE-device measured at 150 mV, respectively. Reprinted from Appl. Phys. Lett. 109, 232402 (2016) and J. Appl. Phys. 122, 223904 (2017), with the permission of AIP Publishing.

**A. Thin film deposition by EB evaporation**

The first series of samples were fabricated by EB evaporation. After the cleaning process, the Si substrates were introduced into an ultra-high-vacuum EB evaporation chamber with a base pressure of $8 \times 10^{-6}$ Pa to deposit a 10 nm Fe layer and finally capped with a 3 nm Au thin film. In order to enhance the spin injection efficiency from Fe to Si, we inserted an MgO/Ge double layer between the Fe electrodes and Si substrates. According to the literature, MgO is a promising spin-dependent tunnel barrier for efficient spin injection from a ferromagnetic electrode into semiconductors (SCs) [16,17]. However, it is difficult to grow a very thin MgO layer with a thickness of 1~2 nm on Si at room temperature to be used as a tunnel barrier. On the other hand, epitaxial growth of MgO on Ge has been reported [18,19]. Moreover, deposition of a smooth thin film of Ge on Si has been demonstrated at low deposition temperature [20]. Therefore, we deposited an ultrathin (1 nm) Ge film as a buffer layer between MgO and Si, to improve the quality of the MgO layer deposited at room temperature. Here, we used 2 nm-thick MgO / 1 nm-thick Ge double layers as a tunnel barrier between the Si channel and 10 nm-thick Fe electrodes. The Fe layer was capped with a 3 nm-thick Au layer to prevent oxidation. All of the deposition processes were conducted in the same EB evaporation chamber without breaking the vacuum.

**B. Thin film growth by MBE**

In the second series of samples, in order to improve the crystal quality of the tunnel barrier and the Fe electrodes, we grew the Fe/MgO/Ge stack by MBE. After the cleaning process, the Si substrates were introduced into a MBE chamber with a base pressure of $1 \times 10^{-8}$ Pa for growing the stack, which consisting of (from top to bottom) a 10 nm Fe layer, 2 nm MgO tunnel barrier, 1 nm Ge, and 1.1 nm MgO cap layer. Knudsen cells were used for thermal evaporation of Ge and Fe, while a low-power electron-beam evaporator was used to deposit MgO with a slow rate of 0.03 Å/s. In-situ reflection high energy electron diffraction confirmed that the MgO layer grown on the Ge buffer layer became crystalline when the MgO thickness exceeded 1 nm. Furthermore, X-ray diffraction has been employed to investigate the crystal quality of the tunnel barrier, and has confirmed that the epitaxial relationship is Fe(001) ∥ MgO(001) ∥ Ge(001). Based on these results, we expect that the crystal quality of the MBE-grown MgO/Ge double layer is much better than that grown by EB evaporation in the first series [14].

### C. Nanofabrication of spin-valve devices

After the thin film deposition, we used electron-beam lithography (EBL) and ion-milling techniques to fabricate nanoscale Si spin-valve devices. Fig. 1(a) shows the schematic structure of our devices examined here. First, we used the EBL and lift-off technique to pattern a 30 nm-thick Au hard mask with a 20 nm-long and 100 nm-wide gap in between, followed by Ar ion milling to etch the exposed Fe area and define a 20 nm-long Si channel. Then, the Au (40 nm) / Cr (5 nm) pad electrodes were deposited by EB evaporation. Fig. 1(b) shows a top-view scanning electron microscopy (SEM) image of a fabricated device. A nanoscale Si channel with a length of 20 nm was formed between the Fe electrodes. In the following sections, the spin-valve devices with Fe/MgO/Ge layers deposited by EB (MBE) method are denoted as EB (MBE) spin-valve devices

### CURRENT – VOLTAGE CHARACTERISTICS

We investigated the conductance behavior of our devices by measuring current – voltage characteristics (*I-V* curves). We found non-linear *I-V* in all the devices, suggesting that tunneling transport through the tunnel barrier contact is dominant. Fig. 1(c) shows representative *I-V* curves of a MBE spin-valve device at various temperatures. A strong dependence of *I-V* curves on temperature has been observed as well. At low temperatures, thermionic emission of electrons from the Fe electrodes to the Si over the tunnel barrier is suppressed. In addition, free carriers in the Si channel are partly quenched. Therefore, the current becomes smaller at low temperatures. This suggests that the resistance of the device is dominated by the transport through the spin-valve structure consisting of Fe(interface) / (MgO/Ge) / Si-channel / (Ge/MgO) / Fe(interface). Fig. 1(d) and Fig. 1(e) show the resistance of an EB-device and an MBE-device as a function of temperature, respectively. In both devices, we observed rapid increases in resistance as temperature decreased, from the order of $\sim 10^1$ $\Omega$ at 300 K to $\sim 10^3$ $\Omega$ and $10^8$ $\Omega$ at 15 K, respectively, for the EB-device and the MBE-device. From these data, we can estimate the contribution of the parasitic anisotropic magnetoresistance (AMR) of the Fe electrodes to the total spin-valve effect MR as follows. The total device resistance $R$ can be decomposed into two components: $R = R_{Fe} + R_{sv}$, where $R_{Fe}$ is the parasitic resistance of the Fe electrodes and $R_{sv}$ is the intrinsic resistance of the Fe(interface)/ (MgO/Ge)/ Si/(Ge/MgO)/ Fe(interface) spin-valve structure. Since the parasitic $R_{Fe}(T)$ decreases as temperature decreases, $R_{Fe}(T) \leq R_{Fe}(300K) < R(300K) \approx 30$ $\Omega$. The AMR effect $|\Delta R_{Fe}/R_{Fe}|$ of Fe is in the order of 0.1%, thus $|\Delta R_{Fe}|$ is in the order of 0.3 $\Omega$. From this consideration, we can estimate the contribution of the parasitic AMR effect of $|\Delta R_{Fe}/R(T)|$ at various temperatures. For example, in the MBE-device, $|\Delta R_{Fe}/R(T)|$ at 200 K should be in the order of $10^{-4}$, while that at 15 K should be in the order of $10^{-8}$. Therefore, we conclude that the contribution of the parasitic AMR effect of the Fe electrodes is negligible at low temperatures.

### SPIN-DEPENDENT TRANSPORT CHARACTERISTICS

### A. Local spin-valve effect

In this work, we use the 2-terminal local spin-valve effect to characterize the spin transport. We show that it is possible to distinguish the intrinsic spin-valve effect from parasitic local effects by systematic measurements of the bias voltage dependence, temperature dependence, and magnetic-field direction dependence of the magneto-resistance (MR). Fig. 2 (a) and (b) show the MR characteristics of our spin-valve devices ((a) EB-device, (b) MBE-device) measured at low temperature with a magnetic field applied along the Si channel (*x*-direction in Fig. 1(a)). Blue dots are the resistance data taken when the magnetic field was swept from +3 kG to -3 kG, while red dots are taken when the magnetic field was swept from -3 kG to +3 kG. Fig. 2(a) shows a clear jump of resistance $\Delta R$ up to 12 $\Omega$ (corresponding to $\Delta R/R = 0.8\%$) of the EB-device. For the MBE-device, as shown in Fig. 2(b), we observed a huge drop of resistance $|\Delta R| \sim 57$ kΩ, corresponding to $|\Delta R/R| = 3\%$, which is more than triple that of the EB-device. This result shows the importance of crystal quality for improving the spin-valve effect signal. Note that we observed negative MR at 15 K for the MBE-device, and the sign of its MR will be discussed later. The $|\Delta R|$ and $|\Delta R/R|$ values in the MBE-device are five orders of magnitude larger than that of the parasitic AMR effect of the Fe electrodes at 15 K ($|\Delta R_{Fe}| \sim 0.3$ $\Omega$ and $|\Delta R_{Fe}/R| = 10^{-8}$). Furthermore, we measured $|\Delta R|$ at various bias voltage *V*, and found that $|\Delta R|$ strongly depends on *V*, as shown in Fig. 2(c). The $|\Delta R|$-*V* relationship closely follows *R-V*. These cannot be explained by the AMR effect, because $|\Delta R_{Fe}|$ does not depend on the bias voltage. From the above quantitative and qualitative considerations, we conclude that the contribution of the parasitic AMR effect is negligible, and that the observed MR effect originates from the spin-dependent tunneling process of electrons between the Fe electrodes and the Si channel through the MgO/Ge barrier. In the following sec-tions, we will concentrate on spin-valve effect of the MBE-device.

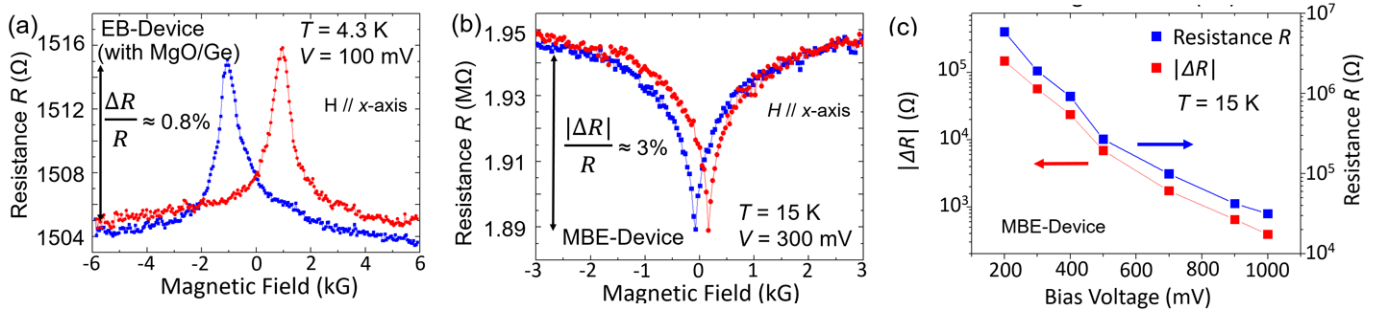

**Fig. 2:** Magnetoresistance characteristics of (a) the EB-device measured at 4.3 K with a bias voltage of 100 mV, and (b) the MBE-device measured at 15 K with a bias voltage of 300 mV. Blue dots are the resistance data taken when the magnetic field was swept from +3 kG to -3 kG, while red dots are taken when the magnetic field was swept from -3 kG to +3 kG. (c) Magnetoresistance |$\Delta R$| and device resistance $R$ of the MBE-device as a function of bias voltage at 15 K. Reprinted from Appl. Phys. Lett. 109, 232402 (2016) and J. Appl. Phys. 122, 223904 (2017), with the permission of AIP Publishing.

### B. Magnetic-field direction dependence

Besides the AMR effect, there is the TAMR effect at the Fe/MgO/Ge/Si interface, which possibly affects the two-terminal voltage as well. This TAMR effect is due to the dependence of the tunneling density of states (DOS) in the FM electrodes on the magnetization direction. In order to distinguish the intrinsic spin-valve effect from the TAMR effect at the Fe/MgO/Ge/Si interface, we have measured the dependence of the MR curve on the mag-neticfield direction $\phi$ with respect to the $x$-direction, as shown in Fig. 3(a). Fig.s 3(b)-3(f) show the MR curves of the MBE-device measured at 100 K when the magnetic field was applied in the film plane along $\phi = 0º$, 30º, 45º, 60º, and 90º, respectively. In the case of TAMR, the resistance state of the FM electrodes depends on the magnetization direction, then the MR curve would be reversed or changed in shape when $\phi$ changes from 0º to 90º [23,24,25]. However, we have observed the same shape and polarity of the MR curves for all the $\phi$ values, indicating that the observed MR depends only on the relative angle between the magnetic moments of the two Fe electrodes. Thus, we conclude that the observed MR is caused by the spin-valve effect.

### C. Temperature dependence

Fig. 4(a) shows the temperature dependence of the spin-valve MR ratio ($\Delta R/R$) of the MBE-device. In this Fig., the negative (positive) spin-valve MR ratio corresponds to the resistance drop (jump) in the MR curves. We found that the local spin-valve MR ratio is negative at tempera-tures lower than 200 K but turns to positive as usual at temperatures higher than 200 K. The inset of Fig. 4(a) shows a representative positive MR curve at 250 K. This phenomenon is unusual and has not been observed be-fore in Si-based spin-valve devices.

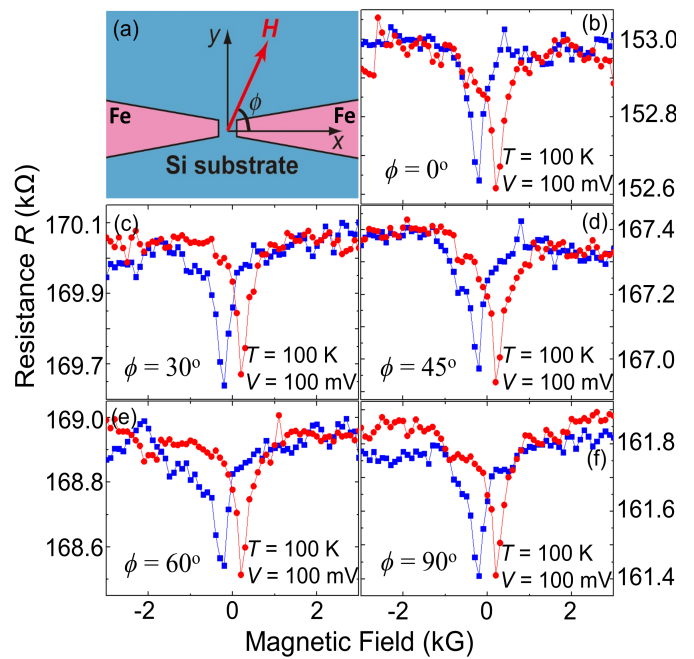

**Fig. 3:** (a) Measurement configuration of the magnetic field direction dependence of the magnetoresistance of the MBE-device. Here, the magnetic field is applied in the $x$-$y$ plane along the angle $\phi$ with respect to the Si channel (the $x$-direction). (b)-(f) Magnetoresistance curves measured at 100 K with a bias voltage of 100 mV, when $\phi = 0º$, 30º, 45º, 60º, and 90º, respectively. Reprinted from J. Appl. Phys. 122, 223904 (2017), with the permission of AIP Publishing.

For understanding the mechanism of this inverse spin spin-valve effect, we focus on the electron tunneling process between the Fe electrodes and Si channel in our devices. Fig. 1(d) shows that the resistance of the MBE-device increases exponentially as temperature decreases, from the order of $10^1$ Ω at 300 K to $10^8$ Ω at 4 K. This strong temperature dependence of the device resistance $R$ suggests that electrons tunnel through the barrier by transport via some defect states in the barrier, whose energy level is much lower than the intrinsic barrier height between Fe an MgO. Such defect states in MgO barriers have been widely studied. For instance, in Fe/MgO/Fe magnetic tunnel junctions, oxygen vacancy defects in the MgO barrier form gap states at about 1.2 eV below the conduction band bottom of MgO, which reduces the MgO barrier height to 0.39 eV [26]. Hence, we assume the existence of such defect-induced gap states inside the MgO barrier of our spin-valve devices. We propose a model using these defect-induced gap states to explain the inverse spin-valve effect, as shown in Fig. 4(b) and (c). In our model, the two Fe electrodes work as a spin injector and detector. Spin-polarized electrons tunnel from the Fe spin injector to the Si channel through such gap states inside the MgO barrier. These electrons then transport in the Si channel without or with little scattering (ballistic or quasi ballistic transport), and reach the opposite MgO barrier with higher kinetic energy. Finally, the electrons tunnel through gap states inside this MgO barrier to the Fe spin detector. In this picture, the relevant gap states are those with energy levels higher than the Fermi level of the Fe spin injector. At high temperatures, many electrons with high enough thermal energy in the Fe spin injector can tunnel to the gap states (thermally activated tunneling), and the device resistance is low (~ $10^1$ Ω at 300 K). However, at low temperatures, the number of electrons that have enough thermal energy rapidly decreases, resulting in much higher resistance (~ $10^8$ Ω at 4 K). Then, only a limited number of the gap states whose energy levels are very close to the Fermi level of the Fe spin injector can allow such tunneling. If these gap states are filled with majority spins whose residence time is long enough at low temperatures, only minority spins from the spin injector can pass through. This is a so-called "spin-blockade", a phenomenon originating from the Pauli exclusion principle, and has been observed in several quantum dot systems as well as defect states in semiconductors [27,28]. On the contrary, the electrons that arrive at the opposite MgO barrier can have higher kinetic energy, thus they can tunnel through many available gap states whose energy levels are higher than the Fermi level of the Fe spin detector. Hence, spin-blockade strongly occurs mostly at the spin injector side, not at the spin detector side. Therefore, the device resistance becomes high (low) at parallel (antiparallel) magnetization configuration, as shown in Fig. 4(b) and 4(c). This explains the observed inverse spin-valve effect at low temperatures. As temperature increases, there are more available gap states with higher energy levels for thermally activated tunneling from the spin injec-tor. Furthermore, the residence time of electrons at the defect states can be much shorter at high temperatures. For this reason, the spin-blockade is not effective anymore, thus the conventional spin-valve effect is dominant at higher temperatures. Our model is also consistent with the impurity assisted tunneling model proposed for expla-nation of the three-terminal Hanle effect observed in some FM/SC systems [29].

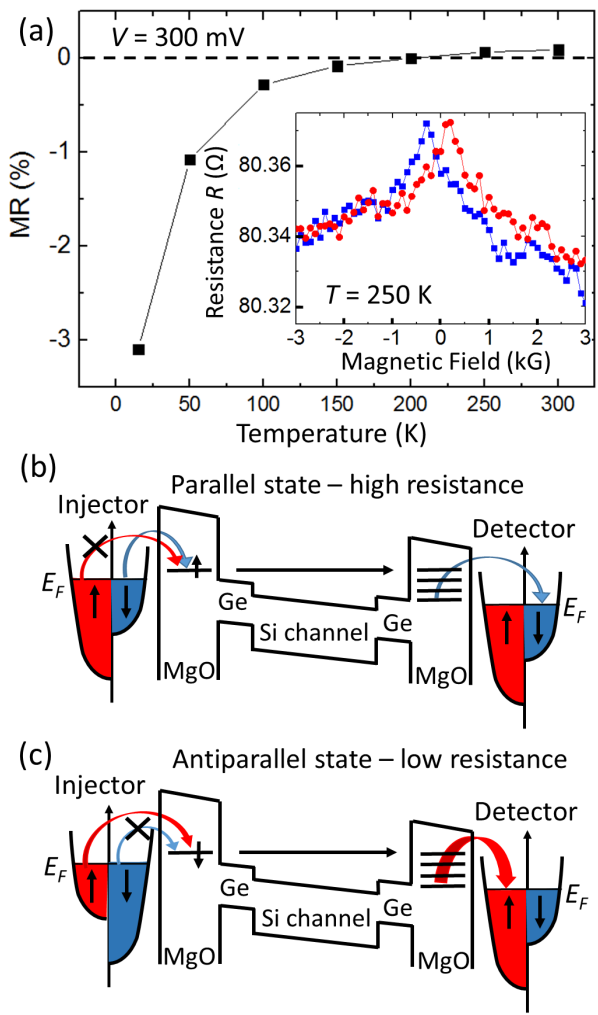

**Fig. 4:** (a) Temperature dependence of the spin-valve MR ratio (defined as $\Delta R/R$) of the MBE-device measured at a bias voltage of 300 mV. Inset shows the MR curve of this device at 250 K. (b)-(c) Proposed spin-dependent transport model through defect-induced gap states in the MgO barriers for parallel and antiparallel magnetization configuration, respectively. At low temperatures, defect-induced gap states in the MgO barrier at the spin injector side are occupied with majority spins, thus only minority spins can tunnel through the MgO barrier because of the Pauli exclusion principle (spin-blockade effect). This leads to the inverse spin-valve effect, where the resistance is high at (b) parallel and low at (c) antiparallel magnetization configuration. Reprinted from J. Appl. Phys. 122, 223904 (2017), with the permission of AIP Publishing.

**D. Spin-dependent output voltage**

In this subsection, we describe the spin-dependent output performance of our devices. The spin-dependent output voltage $\Delta V = (\Delta R/R) \times V$ of the order 100 mV is important for correct read-out in realistic applications. However, previous studies on spin injection into Si channels reported a low read-out voltage of only a few μV in 4 terminal measurements, and about 1 mV in 3 terminal measurements. In this study, by employing the nanoscale Si channel, we have significantly improved $\Delta V$. Figs 5(a) and 5(b) show $\Delta V$ of our EB-device and MBE-device, respectively. In the EB-device, we succeeded in increasing $\Delta V$ up to 13 mV at $V = 1.7$ V. In the MBE-device, we have significantly improved the crystal quality of the Fe/MgO/Ge junctions, as evidenced by the much higher device resistance at 4 K (~ $10^8$ Ω, compared to ~ $10^3$ Ω of our EB-devices). Therefore, we achieved $|\Delta V| = 20$ mV at $V = 0.9$ V, which is the highest value reported so far in lateral Si-based spin-valve devices.

**E. Role of ballistic transport in the nanoscale Si channel**

In this subsection, we describe the advantages of ballistic transport in improving the spin-valve signal in our nanoscale Si channel. Theoretically, in the diffusive transport regime [12,13], the spin polarization of the current at the FM/SC interface is given by $(SP)_I = \left(\dfrac{J_+ - J_-}{J_+ + J_-}\right)_I = \dfrac{\beta}{1 + r_N/r_F}$, in which, $\beta$ is the spin-polarization of the FM layer, $r_F$ and $r_N$ are characteristic resistances defined by $r_F = \rho_F l_{sf}^F$ and $r_N = \rho_N l_{sf}^N$, where $\rho_F$, $\rho_N$ are the resistivity of the FM and SC, and $l_{sf}^F$, $l_{sf}^N$ are the spin-diffusion length in FM and SC. This equation clearly shows the important role of the ratio $r_N/r_F$ in determining the spin polarization at the interface. For example, using $\rho_F \sim 1 \times 10^{-5}$ Ωcm and $l_{sf}^F \sim 2$ nm for Fe, $\rho_N = 2 \times 10^{-2}$ Ωcm for n-Si ($n = 10^{18}$ cm$^{-3}$) and assuming that $l_{sf}^N \sim 10$ μm, the ratio $r_N/r_F$ is ~ $1 \times 10^7$, which is much larger than $\beta$. This means that there is almost no spin polarization at the interface between Fe and n-Si. In case of the FM/SC/FM structure, assuming that the length of the SC layer $t_N \ll l_{sf}^N$, the spin-valve ratio is given by $\dfrac{\Delta R}{R} = 8\beta^2 \left(\dfrac{r_F}{r_N}\dfrac{l_{sf}^N}{t_N}\right)^2$. For the Fe/n-Si interface, even when $t_N = 20$ nm, the ratio $\left(\dfrac{r_F}{r_N}\dfrac{l_{sf}^N}{t_N}\right)^2$ is of the order of $10^{-9}$, which means that there is almost no spin-valve effect. This is so-called the conductivity mismatch problem and can be overcome by introducing a spin-dependent tunnel barrier at the interface between FM and SC [30], which has been widely used to obtain spin injection into SC. However, there are some problems remaining in this method. Even though spin injection from FM into SC through a tunnel barrier has been definitely demonstrated, typical values of the spin-valve ratio reported so far in lateral devices are as small as 0.01% ~ 0.1%. On the other hand, the conductivity mismatch problem can also be overcome by *using ballistic (or quasi-ballistic) transport* in nanoscale SC channels. In the ballis-tic transport regime, Ohm's law and diffusion equations used to derive the spin-polarized current are no longer valid. If the SC channel length is comparable or shorter than the electron mean free path, the transport regime changes from diffusive to (quasi) ballistic. In this case, the transport in the SC channel may be modeled using quantum mechanics instead of classical transport equations of electrons. Then, the transfer matrix method may be used to calculate the spin-polarized transport in the SC channel [2]. Utilizing ballistic transport to overcome the conductivity mismatch problem has obvious advantages such as (i) high spin-valve ratios, similar to tunneling magnetoresistance (TMR), can be achieved; and (ii) the spin-valve effect can be observed even if there is no tunnel barrier, allowing high current driving capability of spin-transistors.

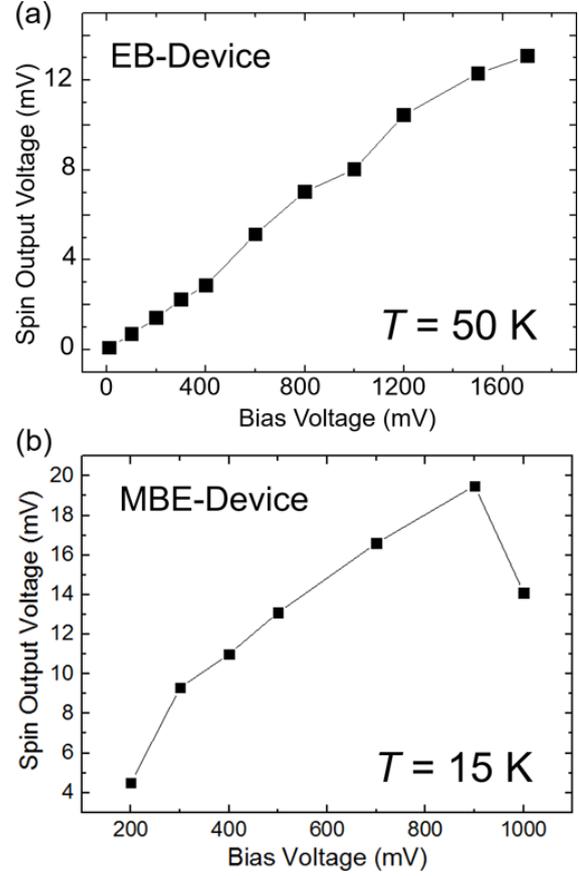

**Fig. 5:** Bias voltage dependence of the spin-dependent output voltage $|\Delta V| = (|\Delta R|/R)V$ of (a) the EB-device, and (b) the MBE-device, respectively. The highest output voltage $|\Delta V|$ of 20 mV was achieved in the MBE-device at the bias voltage of 0.9 V. Reprinted from Appl. Phys. Lett. 109, 232402 (2016) and J. Appl. Phys. 122, 223904 (2017), with the permission of AIP Publishing.

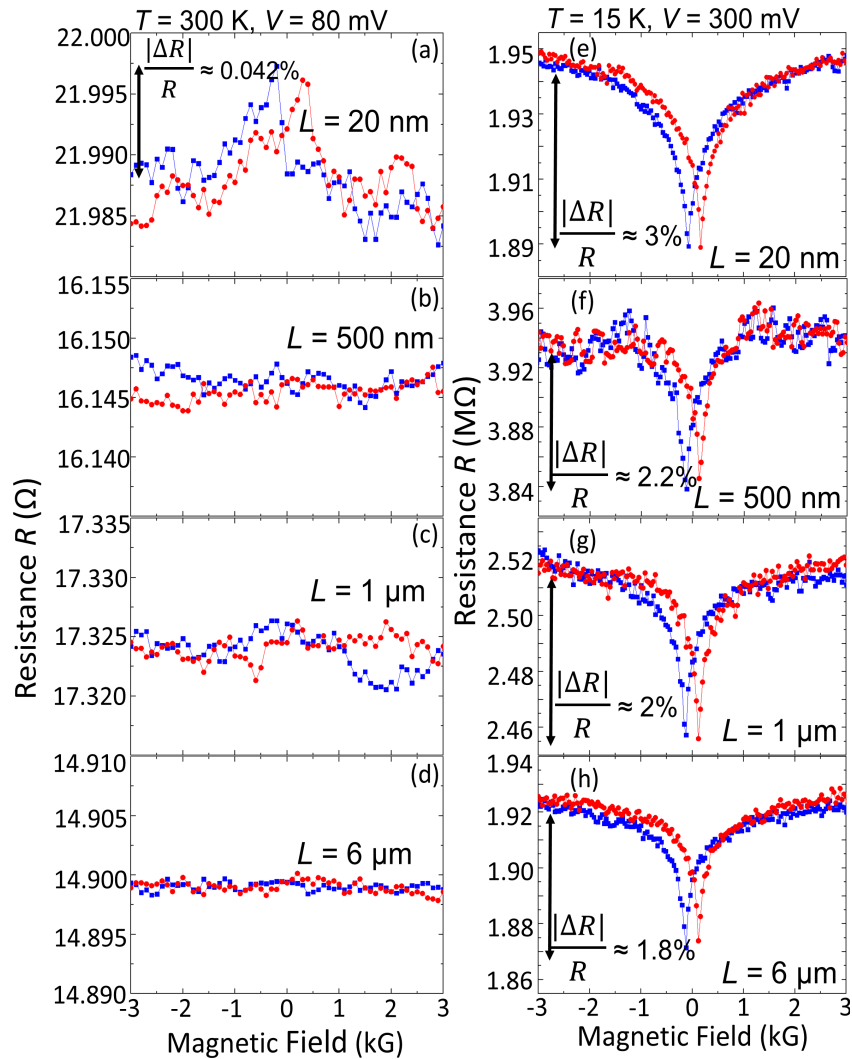

**Fig. 6:** Magnetoresistance of several Si MBE spin-valve devices with different channel length of $L = 20$ nm, 500 nm, 1 μm, and 6 μm, (a)-(d) measured at 300 K with a bias voltage of 80 mV, and (e)-(h) measured at 15 K with a bias voltage of 300 mV. Here, the magnetic field was applied along the current direction. Reprinted from J. Appl. Phys. 122, 223904 (2017), with the permission of AIP Publishing.

In this study, in order to investigate the role of ballistic transport in our nanoscale Si channel, we also prepared and measured Si spin-valve devices with long channel lengths of $L = 500$ nm, 1 μm, and 6 μm by MBE. These lengths are long enough for electrons to transport in diffusive transport regime, as compared with our nanoscale Si spin-valve devices. Fig.s 6 (a) - (d) present the MR curves measured at room temperature for the devices with different channel lengths $L = 20$ nm, 500 nm, 1 µm, and 6 µm, respectively. At room temperature, all the devices show low re-sistance values of the order of 10 Ω. This low resistance indicates the ineffectiveness of tunnel barriers of these devices because of various thermally activated transport processes through defect states in the barriers. Without the barrier resistance, no spin-valve effect can be expected for the long-channel devices. Indeed, the data in Fig. 6(b)-(d) show no spin-valve effect for the devices with $L$ = 500 nm, 1 µm, and 6 µm, as predicted from the diffusion theory. In contrast, Fig. 6(a) shows a clear spin-valve effect of about 0.042% for the nanoscale device, even though there is effectively no barrier for this device at room temperature. This demonstrates the important role of (quasi) ballistic transport in generating the spin-valve effect when there is no tunnel barrier.

Fig.s 6(e)-(f) show the MR curves of these devices at 15 K, measured with a bias voltage of 300 mV. At this low temperature, all the devices showed high resistance values of MΩ, indicating the effectiveness of the barrier in suppressing the thermally activated transport processes through defect states. Such barriers can eliminate the con-ductivity mismatch problem in long-channel devices. As a result, all of the devices showed relatively large spin-valve ratios. However, the data clearly show that the spin-valve ratio systematically decreases with increasing channel length; the MR ratio decreases from ~ 3% for $L$ = 20 nm to 2.2%, 2% and 1.8% for $L$ = 500 nm, 1 µm, and 6 µm, respectively. The drop of the spin-valve ratio is slow, consistent with the long spin-diffusion length in Si.

The results demonstrate two important roles of (quasi) ballistic transport in the nanoscale Si channel; (i) the generation of the spin-valve effect even without a tunnel barrier at room temperature, and (ii) the suppression of spin-flip scattering to achieve a higher spin-valve ratio at low temperature. We have obtained a large spin-valve ratio of 3% in the nanoscale Si channel, which is much larger than those observed in µm-long Si channel devices reported before. The electron transport in our nanoscale devices may be quasi ballistic rather than fully ballistic, because the channel length (20 nm) is not much shorter than that of the mean free path of 20 ~ 40 nm in n-type Si with $n$ = $10^{18}$ cm$^{-3}$ [31]. By using lightly-doped Si substrates and further downsizing the Si channel length to sub-10 nm, we expect to achieve fully ballistic transport and higher spin-valve signals.

## CONCLUSIONS

In this study, we have systematically investigated spin transport in nano-scale Si spin-valve devices. For the MBE-device, a huge spin-valve effect with $|\Delta R|$ up to 57 kΩ, corresponding to $|\Delta R/R|$ = 3%, has been clearly observed. Interestingly, we observed the inverse spin-valve effect at low temperatures, suggesting the possibility of the spin-blockade effect of defect states in the MgO tunnel barrier. The highest spin-dependent output voltage is 20 mV at the bias voltage of 0.9 V at 15 K, which is the highest value reported so far in lateral Si-based spin-valve devices. Our results demonstrate the important role of ballistic transport in improving the spin-valve effect in nanoscale Si spin-valve devices. This work is an important step towards the realization of nanoscale spin-MOSFETs. By using lightly-doped Si substrates and further downsizing the Si channel length to sub-10 nm, we expect to achieve fully ballistic transport and higher spin-valve signals.

**Acknowledgments:** This work was partly supported by Grants-in-Aid for Scientific Research (Grant Nos. 23000010, 16H02095, and 18H01492), Nanotechnology Platform 12025014 (F-17-IT-0010) from MEXT (Ministry of Education, Culture, Sports, Science and Technology), CREST of JST, the Yazaki Foundation, and the Spintronics Research Network of Japan.

*References*
[1] ITRS, Int. Technol. Roadmap Semicond. (2014).
[2] S. Sugahara and M. Tanaka, Appl. Phys. Lett. 84, 2307 (2004).
[3] M. Tanaka and S. Sugahara, IEEE Trans. Electron Devices 54, 961 (2007).
[4] I. Žutić, J. Fabian, and S. Das Sarma, Rev. Mod. Phys. 76, 323 (2004).
[5] V. Zarifis and T.G. Castner, Phys. Rev. B 36, 6198 (1987).
[6] S.P. Dash, S. Sharma, R.S. Patel, M.P. de Jong, and R. Jansen, Nature 462, 491 (2009).
[7] S.P. Dash, S. Sharma, J.C. Le Breton, J. Peiro, H. Jaffrès, J.M. George, A. Lemaître, and R. Jansen, Phys. Rev. B. Phys. Rev. B 84, 054410 (2011).
[8] C.H. Li, O.M.J. Van'T Erve, and B.T. Jonker, Nat. Commun. 2, 245 (2011).
[9] B. Huang, D.J. Monsma, and I. Appelbaum, Phys. Rev. Lett. 99, 177209 (2007).
[10] T. Sasaki, T. Oikawa, T. Suzuki, M. Shiraishi, Y. Suzuki, and K. Tagami, Appl. Phys. Express 2, 053303 (2009).
[11] Y. Aoki, M. Kameno, Y. Ando, E. Shikoh, Y. Suzuki, T. Shinjo, M. Shiraishi, T. Sasaki, T. Oikawa, and T. Suzuki, Phys. Rev. B. 86, 081201 (2012).
[12] G. Schmidt, D. Ferrand, L.W. Molenkamp, A.T. Filip, and B.J. van Wees, Phys. Rev. B 62, R4790 (2000).
[13] A. Fert and H. Jaffrès, Phys. Rev. B 64, 184420 (2001).
[14] D.D. Hiep, M. Tanaka, and P.N. Hai, Appl. Phys. Lett. 109, 232402 (2016).
[15] D.D. Hiep, M. Tanaka, and P.N. Hai, J. Appl. Phys. 122, 223904 (2017).
[16] X. Jiang, R. Wang, R.M. Shelby, R.M. Macfarlane, S.R. Bank, J.S. Harris, and S.S.P. Parkin, Phys. Rev. Lett. 94, 056601 (2005).

[17] C. Martínez Boubeta, E. Navarro, A. Cebollada, F. Briones, F. Peiró, and A. Cornet, J. Cryst. Growth 226, 223 (2001).
[18] K.-R. Jeon, C.-Y. Park, and S.-C. Shin, Cryst. Growth Des. 10, 1346 (2010).
[19] W. Han, Y. Zhou, Y. Wang, Y. Li, J.J.I. Wong, K. Pi, A.G. Swartz, K.M. McCreary, F. Xiu, K.L. Wang, J. Zou, and R.K. Kawakami, J. Cryst. Growth 312, 44 (2009).
[20] C. Schöllhorn, M. Oehme, M. Bauer, and E. Kasper, Thin Solid Films 336, 109 (1998).
[21] X. Lou, C. Adelmann, S.A. Crooker, E.S. Garlid, J. Zhang, K.S.M. Reddy, S.D. Flexner, C.J. Palmstrøm, and P.A. Crowell, Nat. Phys. 3, 197 (2007).
[22] R. Nakane, S. Sato, S. Kokutani, and M. Tanaka, IEEE Magn. Lett. 3, 3000404 (2012).
[23] C. Gould, C. Rüster, T. Jungwirth, E. Girgis, G.M. Schott, R. Giraud, K. Brunner, G. Schmidt, and L.W. Molenkamp, Phys. Rev. Lett. 93, 117203 (2004).
[24] C. Rüster, C. Gould, T. Jungwirth, J. Sinova, G.M. Schott, R. Giraud, K. Brunner, G. Schmidt, and L.W. Molenkamp, Phys. Rev. Lett. 94, 27203 (2005).
[25] J. Moser, A. Matos-Abiague, D. Schuh, W. Wegscheider, J. Fabian, and D. Weiss, Phys. Rev. Lett. 99, 056601 (2007).
[26] S. Yuasa, T. Nagahama, A. Fukushima, Y. Suzuki, and K. Ando, Nat. Mater. 3, 868 (2004).
[27] K. Ono, D. G. Austing, Y. Tokura, and S. Tarucha, Science 297, 1313 (2002).
[28] B. Weber, Y.H.M. Tan, S. Mahapatra, T.F. Watson, H. Ryu, R. Rahman, L.C.L. Hollenberg, G. Klimeck, and M.Y. Simmons, Nat. Nanotechnol. 9, 430 (2014).
[29] O. Txoperena, Y. Song, L. Qing, M. Gobbi, L.E. Hueso, H. Dery, and F. Casanova, Phys. Rev. Lett. 113, 146601 (2014).
[30] E.I. Rashba, Phys. Rev. B 62, R16267 (2000).
[31] B. Qiu, Z. Tian, A. Vallabhaneni, B. Liao, J.M. Mendoza, O.D. Restrepo, X. Ruan, and G. Chen, Europhys. Lett. 109, 57006 (2015).

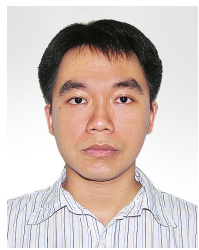

**Duong Dinh Hiep** received his B.E. in physics from Ho Chi Minh City University of Science in 2008, and his M.E. in material science from Japan Institute of Advance Science and Technology in 2015. He is currently a PhD candidate at the Department of Electrical and Electronic Engineering, Tokyo Institute of Technology. His research focuses on spin transport in silicon-based nano-scale spintronics devices.

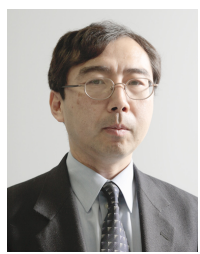

**Masaaki Tanaka** is the director of the Center for Spintronics Research Network (CSRN), and a professor at the Department of Electrical Engineering & Information Systems, Graduate School of Engineering, The University of Tokyo. He received his B.E., M.E., and PhD degrees in electronic engineering from the University of Tokyo, Japan, in 1984, 1986, and 1989, respectively. His research field is electronic materials and device applications, in particular, spintronics materials, spin-related phenomena, and devices including magnetic semiconductors, ferromagnet/semiconductor heterostructures and nanostructures, magnetic tunnel junctions, spin transistors and other devices. He has authored and coauthored over 250 scientific publications, and presented over 120 invited talks at international conferences and meetings.

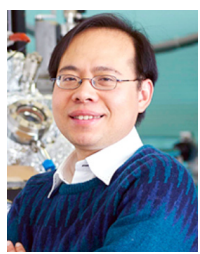

**Pham Nam Hai** is an associate professor at the Department of Electrical and Electronic Engineering, Tokyo Institute of Technology, and a visiting associate professor at the Center for Spintronics Research Network (CSRN), the University of Tokyo. He received B.E., M.E. and PhD degrees in electronic engineering from the University of Tokyo, Japan, in 2004, 2006, and 2009, respectively. His research interests include ferromagnetic nanoclusters, ferromagnetic semiconductors, ferromagnet/semiconductor hybrid systems, topological insulators, and their applications to spintronic devices.